\documentstyle[epsfig]{mn}
\begin{document}

\title
[Lensing by Spiral Galaxies] 
{Strong gravitational lensing by spiral galaxies} 
\author
[Ole M\"oller and A.\,W. Blain]
{
Ole M\"oller \& A.\,W. Blain\\
Cavendish Laboratory, Madingley Road, Cambridge, CB3 OHE.\\
}
\maketitle

\begin{abstract}
We investigate gravitational lensing by a realistic model of disk galaxies. Most of 
the mass is contained in a large spherical isothermal dark matter halo, but the 
potential is modified significantly in the core by a gravitationally-dominant 
exponential disk. The method used is adapted from a very general multi-lens 
ray-tracing technique developed by M\"oller. We investigate the effects of the 
disk-to-halo mass ratio, the disk scale length, the disk inclination to the line of sight 
and the lens redshift 
on two strong-lensing cross sections: the cross section for multiple imaging, and 
the cross section for large magnifications, in excess of a factor of 10. We find that the
multiple-imaging cross section can be enhanced significantly by an almost edge-on Milky Way disk as compared with a singular isothermal 
sphere (SIS) in individual cases; however, when averaged over all disk inclinations, the cross section 
is only increased by about fifty per cent. These results are consistent with other 
recent work. The presence of a disk, however,
increases the inclination-averaged high-magnification cross section by an order of 
magnitude as compared with a SIS. This result has important implications 
for magnification bias in future lens surveys, particularly those in the 
submillimetre waveband, where dust extinction in the lensing galaxy has no 
effect on the brightness of the images.
\end{abstract}  


\section{Introduction}
There is ever increasing interest in gravitational lensing by individual galaxies. Deep 
surveys in the radio waveband, for example the CLASS survey (Myers et al. 1995), 
have identified many galaxy lensed systems.
In many cases, models of the lensing mass distribution as spherical or elliptical 
isothermal halos have been reasonably successful in reproducing the observed 
image positions and magnification ratios (for example, Rhee 1991). An 
increasing number of observed lens systems have, however, image configurations that are 
not consistent with this simple picture (for example Jaunsen \& Hjorth 
1997). These have recently lead to the consideration of lens models that include a 
disk component within a dark halo (Bartelmann \& Loeb 1998; Keeton \& 
Kochanek 1998; Koopmans, de Bruin \& Jackson 1998; Maller, Flores \& Primack 1997; Wang \& Turner 1997).\\
Since a disk with mass comparable to that of the halo would be gravitationally 
unstable, the mass contained in the disk is not expected to exceed more 
than about 10 per cent of the halo mass. Even so, within a few kpc of the core
of such a galaxy the projected surface mass density on the lens plane is dominated 
by the disk component and this enhancement should lead to an increase in the strong-lensing cross section.
The effect is expected to be particularly significant for a nearly edge-on disk. \\
Wang \& Turner (1997) calculated the cross sections for multiple imaging for 
lenses consisting of a singular isothermal sphere (SIS) and a thin uniform disk, 
and found a large increase at large disk inclinations; however, the model of a
uniform galaxy disk is not very realistic. Maller et al. (1997) used an exponential 
disk mass distribution to fit the two-image lens system B\,1600+434 
(Jackson et al. 1995) with reasonable success. 
Keeton \& Kochanek (1998) presented a model that combined three different 
elliptical mass components to produce an analytical approximation to an exponential disk potential with a flat rotation curve at large distances. 
This model involves many parameters and the potential differs significantly from that of
an exponential profile near the core of the galaxy.
While we were preparing this paper, Bartelmann \& Loeb (1998) discussed the 
statistical significance of lensing by spiral disks using a technique similar to 
ours to derive strong lensing optical depths, including the effect of dust 
extinction along the lines of sight to different images. We consider these effects in detail in a subsequent paper (Blain, M\"oller \& Maller; in preparation).

This work was motivated by interest in the effects of disk lensing on the 
population of lenses in the millimetre/submillimetre waveband, where, in some 
circumstances, magnification biases are expected to be very large and up to 
5 to 10 per cent of sources detected in a survey could be lensed (Blain 1996, 1997,
1998). An enhanced strong-lensing cross section due to disks could increase
this effect even further (Blain et al. in preparation), especially because dust extinction is unimportant in the submillimetre waveband. 

We adapted an existing general ray-tracing method, which was first developed 
to study weak lensing (M\"oller 1997), to solve the problem. The method and 
lens model are outlined in Section\,2. We then investigate the effects of the 
surface density and scale length of the disk on the cross section for the
formation of multiple images, which is essential for determining whether a 
source in a survey will be classified as a lens in Section 3, and the cross section for large
magnifications, which determines the importance of magnification bias (Borgeest,
Linde \& Refsdal 1991) in Section\,4. In Section\,5 we
compare our results with those of previous authors and briefly discuss their 
implications for lensing statistics and magnification biases. We assume that 
the density parameter $\Omega_0=1$ and Hubble's constant 
$H_0=60\,\mathrm{km\,s^{-1}\,Mpc^{-1}}$. Angular diameter distances are 
calculated assuming a smooth distribution of matter.

\begin{figure}
\label{rotation}
\begin{center}
\epsfig{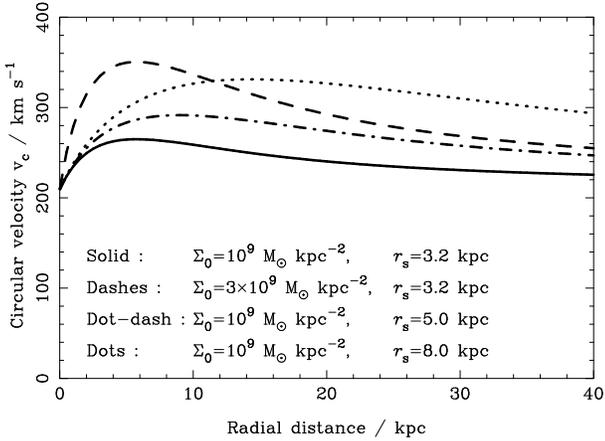}
\end{center}
\caption{Rotation curves for an exponential disk in an SIS halo with 
$\sigma_{\mathrm{v}}=150\,\mathrm{km\,s^{-1}}$. The circular velocity $v_{\mathrm{c}}=\sqrt{2}\sigma_{\mathrm{v}}$.}
\end{figure}

\section{Method} 

Even for a uniform disk, the analytical derivation of the form of caustics and 
critical lines is very involved. In order to study lensing by exponential disks, we 
adapted an existing general multi-lens-plane ray-tracing routine (M\"oller 1997), 
in which a regular grid of $R_{\mathrm{I}}\times R_{\mathrm{I}}$ images is 
mapped onto the source plane by propagating light rays through a set of $N$ 
lens planes. In the present computations, $N=1$. The simulated source plane is finite, and so 
care was taken that no rays are deflected beyond its boundaries. 
In principle, extended sources can be treated, but all the results are derived here 
for point sources; the brightest regions observed in far-infrared-luminous
galaxies at low redshifts are typically less than a few hundred parsecs across (Solomon et al.\ 
1997), corresponding to an angular size of order $10^{-2}$\,arcsec at redshift
$z=2$, and so this approximation is unlikely to affect our applications significantly.
To generate maps of the total magnification on the source plane, we 
find all the images of a regular grid of $R_{\mathrm{S}} \times R_{\mathrm{S}}$ 
point sources using the `grid search method' described by Schneider, Ehlers 
\& Falco (1992). Once all the images of each source are located, we find its total 
magnification by summing all the individual magnifications. 

The accuracy of this procedure depends on the resolution of the grid on the source and image 
planes. If the image resolution is too coarse, then the grid search method 
may miss images near caustics, and so the total magnification is 
underestimated. This problem can be partially overcome by making 
$R_{\mathrm{I}} \gg R_{\mathrm{S}}$. There is a trade-off between the accuracy 
of magnifications near caustics, the coarseness of the resulting map and the
computing time. The cross sections for different magnifications are 
determined by adding together the areas of all pixels in the source plane that 
contain sources with that magnification. Hence, a small value of 
$R_{\mathrm{S}}$ leads to an overestimate of the high-magnification cross 
section as the \emph{average} magnification over the whole pixel area is not 
sampled adequately. To verify the accuracy of our procedure and to find 
optimum values for $R_{\mathrm{I}}$ and $R_{\mathrm{S}}$, we tested its
predictions against analytic results derived for a SIS lens. For resolution ratios 
$R_{\mathrm{I}}/R_{\mathrm{S}} > 2$ the results agreed with the analytical 
predictions to within about 5 per cent. In all computations, $R_{\mathrm{I}}=1000$ and $R_{\mathrm{S}}=400$.

The deflection angle $\mbox{\boldmath $\alpha$}$ is calculated by adding together the 
deflection due to the two different lens components, a SIS halo and a disk,
\begin{equation}
\mbox{\boldmath $\alpha$}=\mbox{\boldmath $\alpha$}_{\mathrm{SIS}}+\mbox{\boldmath $\alpha$}_{\mathrm{disk}}.
\end{equation}
The deflection due to the halo,
\begin{equation}
\mbox{\boldmath $\alpha$}_{\mathrm{SIS}}=\frac{4GM_0}{rR_0c^2}\mbox{\boldmath $r$},
\end{equation}
where $M_0$ is the mass contained within radius $R_0$ and $r$ is the impact parameter. The projection of an inclined thin disk onto the lens plane produces an elliptical surface mass density 
profile. We evaluate $\mbox{\boldmath $\alpha$}_{\mathrm{disk}}$ numerically using a useful 
elliptical surface potential formalism developed by Schramm (1990). 
  
\begin{figure*}
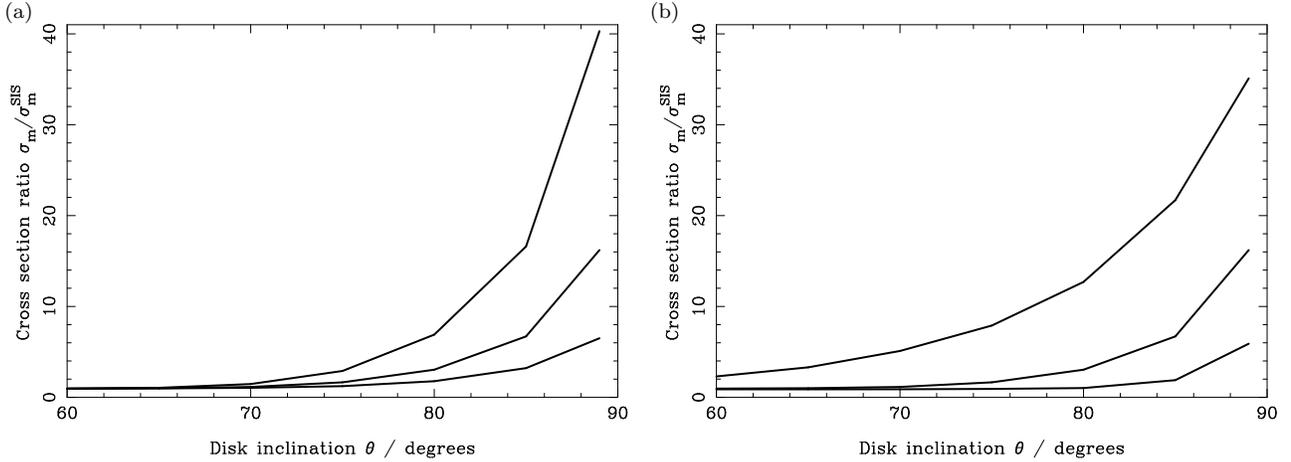

\label{multiplot}
\begin{minipage}{170mm}
(a) \hskip 81mm (b)
\begin{center}
\vskip -5mm
\epsfig{file=fig2a.ps,width=5.8cm,angle=-90} 
\hskip 3mm
\epsfig{file=fig2b.ps,width=5.8cm,angle=-90}
\end{center}
\caption{Multiple-imaging cross sections for a lens consisting of a SIS halo, 
with a velocity dispersion $\sigma_{\mathrm{v}}=150\,\mathrm{km\,s^{-1}}$, and 
an exponential disk with a central surface density $\Sigma_0$ and radial scale
length $r_{\mathrm{s}}$. In (a) $\Sigma_0$ is fixed at 
$10^9\mathrm{M_{\odot}}\,\mathrm{kpc^{-2}}$: the upper, middle and lower
curves correspond to $r_{\mathrm{s}} = 8.0$, 5.0 and $3.2\,\mathrm{kpc}$ 
respectively. In (b) $r_{\mathrm{s}}$ is fixed at $5.0\,\mathrm{kpc}$; the upper, 
middle and lower curves correspond to 
$\Sigma_0=2.6 \times 10^9 \mathrm{M_{\odot}}\,\mathrm{kpc^{-2}}$, 
$1.0 \times 10^9\mathrm{M_{\odot}}\,\mathrm{kpc^{-2}}$ and 
$0.4 \times 10^9\mathrm{M_{\odot}}\,\mathrm{kpc^{-2}}$ respectively. In all cases 
$z_{\mathrm{l}}=0.2$, $z_{\mathrm{s}}=1.5$.}
\end{minipage} 
\end{figure*}

We consider a single model lens at 
redshift $z_{\mathrm{l}}$. 
The lens consists of a SIS dark matter halo with velocity dispersion 
$\sigma_{\mathrm{v}}$ and a disk with a surface mass profile 
$\Sigma(r)=\Sigma_0\exp\left(-r/r_{\mathrm{s}}\right)$, in which 
$r_{\mathrm{s}}$ is the radial scale length. For clarity, no bulge component is 
included, but our model can incorporate additional mass components without 
difficulty. The effect of including a bulge is discussed briefly in Section 4.  
In Fig.\,\ref{rotation} four realistic flat rotation curves predicted by the mass models used in this 
paper are shown. Values of $\sigma_{\mathrm{v}}=150\,\mathrm{km\,s^{-1}}$, $\Sigma_0=10^9\mathrm{M_{\odot}}\,\mathrm{kpc^{-2}}$ and 
$r_{\mathrm{s}}=3.2\,\mathrm{kpc}$ produce a reasonable fit to the rotation curve
of the Milky Way (Binney \& Tremaine 1987).

\section{Multiple-imaging cross section}

For a SIS lens, the cross section for multiple imaging, 
\begin{equation}
\sigma^{\mathrm{SIS}}_{\mathrm{m}}=\left[ 8 \frac{\sigma_{\mathrm{v}}^2}{c^2} D_{LS} \right]^2 \pi,
\end{equation}
(Schneider et al. 1992); $D_{\rm LS}$ is the lens--source angular diameter 
distance. If the lens redshift $z_{\mathrm{l}}=0.2$, the source redshift $z_{\mathrm{s}}=1.5$ and $\sigma_{\mathrm{v}}=150\,\mathrm{km\,s^{-1}}$ then $\sigma^{\mathrm{SIS}}_{\mathrm{m}}=15.1\,\mathrm{kpc^2}$.
Including a disk component increases $\sigma_{\mathrm{m}}$ if the inclination angle of the 
disk $\theta$ exceeds a critical angle $\theta_{\mathrm{c}}$ (equation 10 in Maller et~al. 1997), 
$\theta=0$ and 90\,deg corresponding to a face-on and edge-on disk
respectively. 
Note that we \emph{add} the disk to a halo with a fixed velocity dispersion and consequently, our pure SIS model does not take into account the disk mass. We use this approach because we aim to demonstrate how a disk can increase the lensing cross section without requiring a more massive halo. For a Milky Way type galaxy, the disk-to-halo mass ratio within a ten kiloparsec radius is about 1:2, and including the disk mass into a pure SIS model would increase the velocity dispersion of the halo above the observed value.   

\begin{figure*}
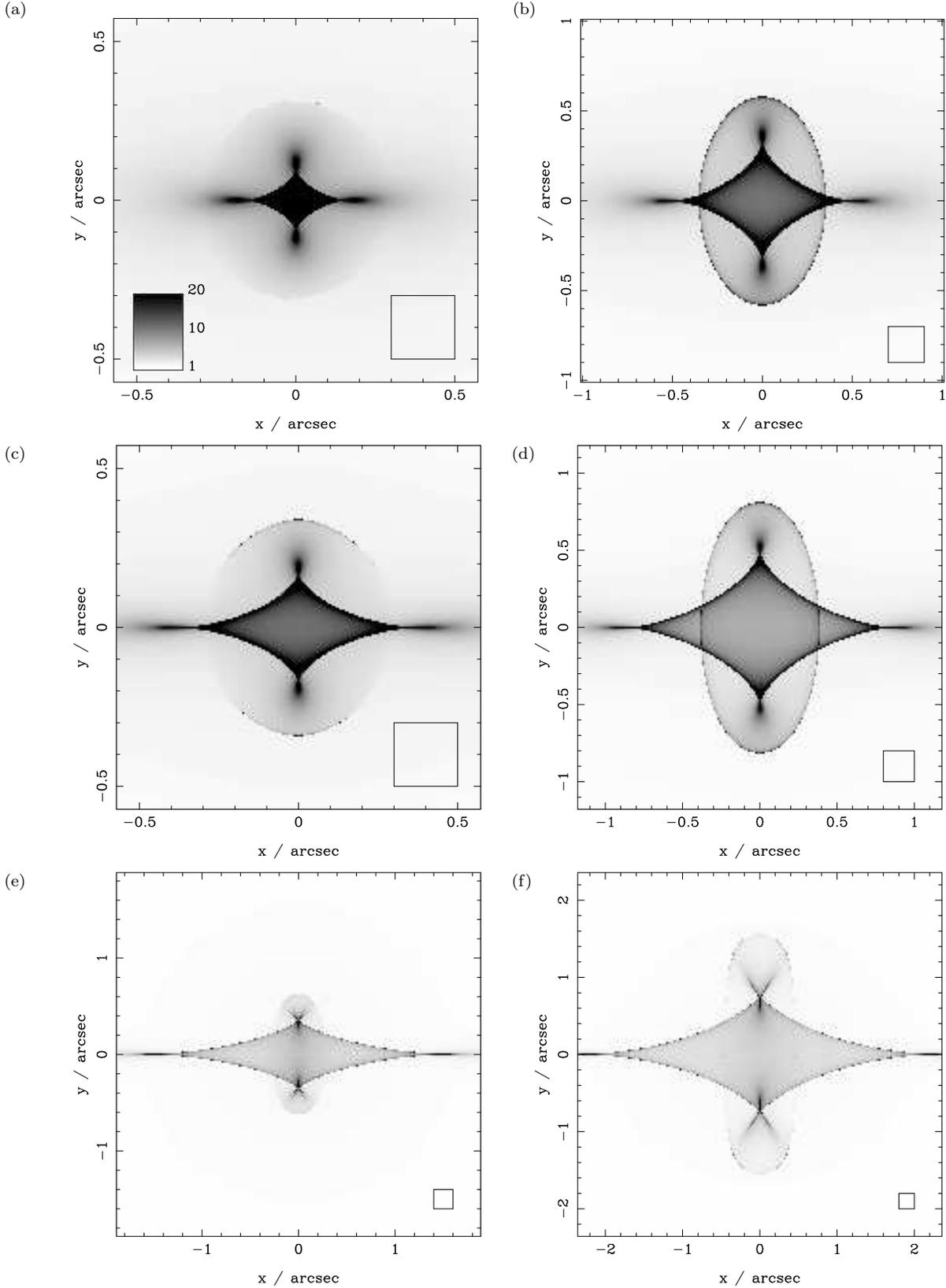

\label{maps}
\begin{minipage}{170mm}
(a) \hskip 81mm (b)
\begin{center}
\vskip -5mm
\epsfig{file=fig3a.ps,width=7.0cm,angle=-90} \hskip 9mm
\epsfig{file=fig3b.ps,width=7.0cm,angle=-90}\\
\end{center}
(c) \hskip 81mm (d)
\begin{center}
\vskip -8mm  
\epsfig{file=fig3c.ps,width=7.0cm,angle=-90}\hskip 9mm
\epsfig{file=fig3d.ps,width=7.0cm,angle=-90}\\
\end{center}
(e) \hskip 81mm (f)
\begin{center}
\vskip -8mm  
\epsfig{file=fig3e.ps,width=7.0cm,angle=-90}\hskip 9mm
\epsfig{file=fig3f.ps,width=7.0cm,angle=-90}\\
\end{center}
\caption{Magnification maps for a lens centred at the origin that consists of an 
SIS halo with $\sigma_{\mathrm{v}}=150\,\mathrm{km\,s^{-1}}$ and a disk 
with $r_{\mathrm{s}}=3.2\,\mathrm{kpc}$ with its major axis lying along the x-axis; 
$z_{\mathrm{l}}=0.2$ and $z_{\mathrm{s}}=1.5$. The grey-scale represents the total magnification of a source located at $x,y$ in the source plane.  In (a), (c) and (e) 
$\Sigma_0=10^9\,\mathrm{M_{\odot}}\,\mathrm{kpc^{-2}}$. In (b), (d) and (f) 
$\Sigma_0=3\times 10^9\,\mathrm{M_{\odot}}\,\mathrm{kpc^{-2}}$. 
$\theta=60$\,deg in (a) and (b), $\theta=70$\,deg in (c) and (d) and 
$\theta=85$\,deg in (e) and (f). The outlined box in the bottom-right corner of each panel encloses a constant area and is shown to indicate the different scales.}
\end{minipage} 
\end{figure*}

The variation of $\sigma_{\mathrm{m}}$ as a function of inclination is shown in Fig.\,\ref{multiplot} for various disk parameters. In all cases, 
$\sigma_{\mathrm{m}}$ increases monotonically with increasing inclination, and 
the effect is considerably more significant for more massive disks. Although the 
parameters are chosen so that the upper curves in Figs.\,\ref{multiplot}(a) \& (b) 
are derived for disks of the same total mass, the predicted cross sections differ significantly at 
small inclinations: the disk with the larger surface mass density has the 
more significant effect. These results agree with the expected dependence of 
$\theta_{\mathrm{c}}$ on the surface mass density reported by Maller et al. 
(1997). The effect on the cross section ratio
 of changing $z_{\mathrm{l}}$ is negligible, as compared with the dependence on the surface mass density and scale length.

The multiple-imaging cross section is increased significantly at large inclinations, 
but if $r_{\mathrm{s}}=3.2\,\mathrm{kpc}$ and 
$\Sigma_0=10^9\,\mathrm{M_{\odot}}\,\mathrm{kpc^{-2}}$, the 
inclination-averaged multiple-lensing cross section, 
\begin{equation}
\bar{\sigma}_{\rm m} =\int_0^{\pi/2} \sin\theta\,\sigma_{\rm m} 
(\theta)\,{\rm d}\theta,
\end{equation}
only exceeds that of the SIS halo by about 50 per cent. Hence,
although high-inclination disks can increase the multiple-lensing cross section 
significantly as compared with a SIS, the inclination-averaged effect of disks is 
small: the effect of a disk can be very important when modelling the image 
configurations in individual lens systems, but the presence of a population of
disks should not have a significant effect on the number of lenses that 
will be identified in a large sample of candidate sources. 
We also investigated the relative number of images. For a Milky Way type galaxy we find that the inclination averaged cross section for the formation of double and triple images is approximately equal, with about 40-50 per cent of the total multiple image cross section being provided by each. Only about 10 per cent of multiply imaged sources will be split into four images. The cross section for a source being lensed into five images is zero within the numerical error. Note that these results do not take into account magnification bias.

\section{High-magnification cross section}

For a SIS lens, the cross section for magnifications $\mu>A$, 
\begin{equation}
\sigma_\mu^{\mathrm{SIS}} (A) = 
\left\{ \begin{array}{l} 4\sigma^{\mathrm{SIS}}_{\mathrm{m}}A^{-2}, 
\quad\mbox{for } A>2; \\ \sigma^{\mathrm{SIS}}_{\mathrm{m}} (A-1)^{-2}, 
\quad\mbox{for } 1<A\leq 2, \end{array} \right. 
\end{equation} 
as can easily be obtained from the magnification curve (Schneider et al. 1992). 
For large magnifications, however, we predict that $\sigma_\mu$ is much larger for a 
SIS plus disk lens as compared with a SIS alone. Magnification maps derived at 
inclinations of 60, 70 and 85\,deg for a lens similar to the 
Milky Way are presented in Fig.\,\ref{maps}, and clearly show the distinctive 
shape of the caustic, with an elliptical component due to the halo and a diamond shaped component, the ``astroid'', due to the disk. For large 
inclinations, the size of the astroid increases significantly, leading to the 
increase of the multiple-imaging cross section $\sigma_{\rm m}$ discussed 
above.

The cross section ratio $\sigma_\mu/\sigma_\mu^{\mathrm{SIS}}$ for 
magnifications greater than A for a galaxy lens similar to the Milky Way are presented in Fig.\,4. The cross sections are derived by adding up the area of all pixels on the source plane with magnifications greater than A. 
The calculated ratios become less smooth at large magnifications because the 
number of high-magnification pixels on the source plane is relatively small; only about 20 pixels in the source plane have magnifications larger than $A=80$.
The magnitude of the fluctuations in the ratios can be used as an estimate of the  
uncertainty in the results of these computations. A large increase in the cross section ratio 
is predicted for inclinations that exceed 60\,deg, at magnifications greater 
than a certain threshold. This threshold magnification is a strong function of 
inclination, and is reduced as the inclination increases. The cross section ratios shown 
in Figs.\,4 \& 5 also have a characteristic peak beyond this threshold magnification. Hence, although the
largest cross section ratios are predicted for almost edge-on disks at magnifications 
greater than 50, at magnifications of about 10 a disk at $\theta = 70$\,deg is
expected to produce a larger cross section than a disk at 90\,deg. 
Hence, nearly edge-on disks are \emph{not necessarily} more effective in producing large magnifications
as compared with disks at smaller inclinations. This is because points interior 
to the astroid caustic lie closer to caustic lines and experience larger
average magnifications when the caustic is smaller, as it is at moderate inclinations.
This effect is clearly demonstrated in Fig.\,3; although the area of the astroid caustic is larger in the lower panels, the mean magnification in the enclosed region is smaller. 

\begin{figure}
\label{high1}
\begin{center}
\epsfig{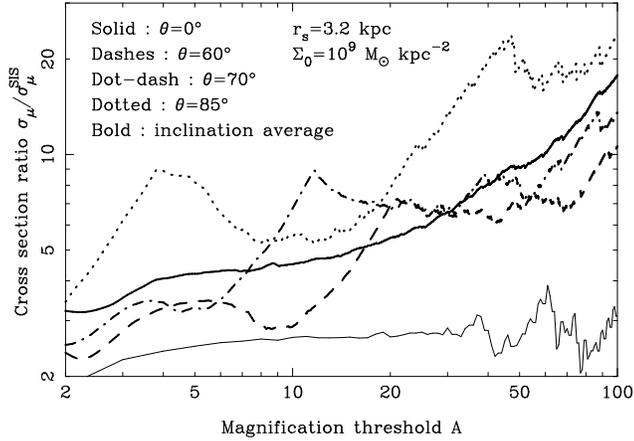}
\end{center}
\caption{High-magnification cross sections for a galaxy with a SIS with 
$\sigma_{\mathrm{v}}=150\,\mathrm{km\,s^{-1}}$ and a Milky Way disk. 
To indicate the uncertainties involved the curves are not smoothed.}
\end{figure}

The inclination-averaged high-magnification cross section
$\bar{\sigma}_\mu$ shown in Fig.\,\ref{high1} was derived by evaluating 
$\sigma_\mu$ at five-degree intervals from 0 to 90\,degrees and approximating 
the integral (equation 4) by a sum. The result is very significant: 
$\bar{\sigma}_\mu$ is greatly increased by a randomly-oriented galaxy 
similar to the Milky Way as compared with a SIS lens, and 
$\bar{\sigma}/\sigma_{\mathrm{SIS}} \approx 10$ at magnifications greater than
about 50. This increase could have a very significant effect on magnification bias in galaxy 
surveys. 

The dependence of $\sigma_{\mu}$ on the scale length and surface density of the disk is shown in 
Fig.\,5. As in the case of the multiple-imaging cross section, the more massive 
disks produce a more significant effect. Again, if the central surface mass 
density is increased, then the relative significance of disks with smaller 
inclinations increases. Adding a central bulge does not have a significant effect on the high-magnification cross section; for a spherical bulge with a $r^{1/4}$-law density profile, extending out to one kiloparsec from the centre and containing a total of 10 per cent of the disk mass, we find that the high-magnification cross section is increased by no more than 50 per cent. The projected surface mass density profile for an elliptical halo is similar to that for an inclined  disk. Thus, we expect that a non-spherical halo which is flattened along the disk is likely to enhance the lensing effect of an inclined disk. However, this effect is not expected to be significant.   

The high-magnification cross section does not depend strongly 
on the redshift of the lens. This suggests that the high-magnification cross section in both, the disk plus SIS halo model and the pure SIS halo model scales with the angular diamater distance in a similar way. Thus, we do not expect $\bar{\sigma}_{\mu}/\sigma^{\mathrm{SIS}}_{\mu}$ to be strongly dependent on the value of the density parameter $\Omega_0$ or the comological constant $\lambda$.

\begin{figure*}
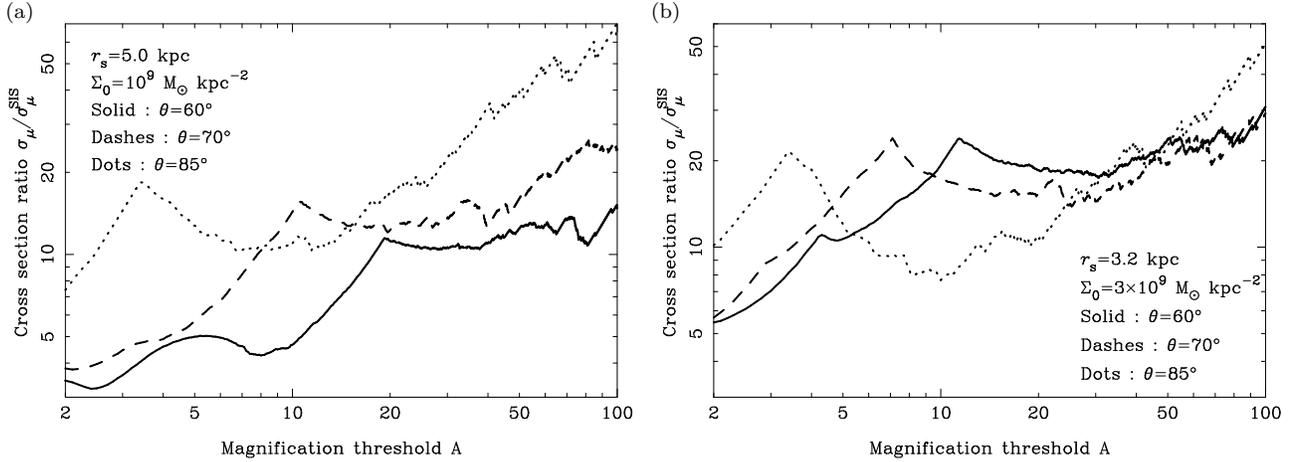

\label{high2}
\begin{minipage}{170mm}
(a) \hskip 81mm (b)
\begin{center}
\vskip -5mm
\epsfig{file=fig5a.ps,width=5.8cm,angle=-90}
\hskip 2mm
\epsfig{file=fig5b.ps,width=5.8cm,angle=-90}
\end{center}
\caption{High-magnification cross sections for galaxy models with different 
disk parameters from those assumed in Fig.\,4. In (a) and (b) the mass of the 
disk is 2.5 and 3 times larger as compared with the model in Fig.\,4.}
\end{minipage}
\end{figure*}

\section{Discussion}

We have studied the effects of lensing by an exponential disk inside a SIS on two
important parameters; the multiple-imaging cross section $\sigma_{\rm m}$, and 
the high-magnification cross-section $\sigma_\mu(A)$. 
The lens model we use has a 
density profile similar to that of the Milky Way. Both $\sigma_{\rm m}$ and 
$\sigma_\mu(A)$ are sensitive to the lens parameters, in particular to the 
inclination angle and the disk-to-halo mass ratio.

The results indicate that disks, and especially almost edge-on disks, have a 
significant effect on the image configurations produced by individual lenses, 
which agrees with the results of previous work by Maller et al. (1997) and Wang \& Turner 
(1997), but that the statistical effect on $\sigma_{\rm m}$ averaged over all disk inclinations is unlikely to be very 
important. 
  
The more significant result is that including an exponential disk at an inclination 
greater than about 60\,deg increases $\sigma_\mu$ at $A>10$ by an order of 
magnitude as compared with an SIS lens, and that the inclination-averaged cross 
section $\bar \sigma_\mu$ is increased by a similar factor. Hence, disks
could have a large effect on the predicted abundances of 
strongly magnified galaxy lenses. This implies that in surveys for 
which dust extinction is unimportant, such as those in the radio and
submillimetre wavebands, the number of detected, strongly lensed sources will increase 
(Blain 1996, 1997), and that they will tend to be 
preferentially associated with edge-on spiral galaxies (Blain et al., in preparation). 
The preliminary results of the CLASS survey suggest that this is indeed the 
case (Browne et al. 1997).
A recent preprint by Bartelmann \& Loeb (1998) 
discussed the statistical significance of lensing by spiral galaxies using a 
technique similar to ours and included the effect of dust extinction in the lensing 
galaxy. They did not discuss the effects of different scale lengths or disk masses 
on the high-magnification cross sections and so it is difficult to compare our 
results in detail, but they appear to be in broad agreement, bearing in mind 
that they assume $\Omega_0=0.3$ and $H_0=50\,\mathrm{km\,s^{-1}\,Mpc^{-1}}$ 
and that their disk model deviates from our exponential profile.

\section{Conclusions}

\begin{enumerate}
\item The configuration of images in an individual lens can be modified
significantly by the presence of a disk component at a large inclination, but 
when averaged over all inclinations the multiple-imaging cross section of the lens is increased by only about 50 per cent. 
\item The large-magnification cross section is increased considerably by including a disk component in the lens model. For a Milky Way disk the effect is 
very significant at inclinations in excess of a moderate value of about 60\,deg,
and so the inclination-averaged cross section is also expected to increase 
significantly, by an order of magnitude at magnifications of order 10 as compared with a 
singular isothermal sphere lens.
\end{enumerate}

\section*{Acknowledgements}

We thank Ariyeh Maller for useful and interesting conversations, one of which 
motivated this paper, and Priya Natarajan, Malcolm Longair and an anonymous referee for providing helpful comments.

\end{document}